\documentclass[pre,a4paper,onecolumn]{revtex4}
\usepackage{graphicx}
\usepackage{amsmath}
\usepackage{bbm}
\usepackage{psfrag}
\usepackage{latexsym}

\newcommand{\bq}{\begin{equation}}
\newcommand{\eq}{\end{equation}}
\newcommand{\ba}{\begin{eqnarray}}
\newcommand{\ea}{\end{eqnarray}}

\newcommand{\Tr}{\mathrm{Tr}\,}

\newcommand{\X}{X}
\newcommand{\Y}{Y}

\newcommand{\A}{A}
\newcommand{\B}{B}

\begin{document}

\title{\bf Spectrum of the Product of Independent Random Gaussian Matrices}

\author{Z. Burda$^1$}\email{zdzislaw.burda@uj.edu.pl}

\author{R.A. Janik$^1$}\email{romuald@th.if.uj.edu.pl}

\author{B. Waclaw$^2$}\email{bwaclaw@staffmail.ed.ac.uk}

\affiliation{$^1$\mbox{Marian Smoluchowski Institute of Physics, Jagellonian 
University, Reymonta 4, 30-059 Krak\'ow, Poland}\\
$^2$\mbox{SUPA, School of Physics and Astronomy, University of Edinburgh, 
Mayfield Road, Edinburgh EH9 3JZ, UK}}

\begin{abstract}

We show that the eigenvalue density of a product $\X=\X_1 \X_2 \cdots \X_M$
of $M$ independent $N\times N$ Gaussian random matrices
in the limit $N\rightarrow \infty$ is rotationally symmetric in the complex plane and is given by a simple expression 
$\rho(z,\bar{z}) = \frac{1}{M\pi} \sigma^{-\frac{2}{M}} |z|^{-2+\frac{2}{M}}$ for $|z|\le \sigma$,
and is zero for $|z|> \sigma$. The parameter $\sigma$ corresponds to the radius of the circular support and is related to the amplitude of the Gaussian fluctuations. This form of the eigenvalue density is highly universal. It is identical for products of Gaussian Hermitian, non-Hermitian, real or complex 
random matrices. It does not change even if the matrices in the
product are taken from different Gaussian ensembles.
We present a self-contained derivation of this result using a planar diagrammatic technique. Additionally, we conjecture that this 
distribution also holds for any matrices whose elements are independent, centered random variables with a finite variance or even more generally for matrices which fulfill Pastur-Lindeberg's condition. We provide a numerical evidence supporting this conjecture.

\end{abstract}

\maketitle

\section{Introduction}
Initiated by Wigner more than 50 years ago and developed by Dyson, Mehta and others, Random Matrix Theory (RMT) has been successfully applied to
various problems ranging from fundamental physics (for a comprehensive review see \cite{guhr}) to engineering and financial applications \cite{jpb}. One of the reasons of such a wide applicability is the universality of many results predicted by RMT. Let us take as an example the problem addressed by Wigner, that is how to determine the energy spectrum and level spacing distribution of a many-body quantum system. Due to many degrees of freedom and sophisticated nature of interactions one has to turn to a statistical description. However, in contrast to statistical mechanics where one fixes the Hamiltonian and averages over possible states of the system, Wigner proposed to treat the very Hamiltonian as a random operator, which in turn can be represented as a large random matrix. Relevant properties of such a matrix are determined by symmetries of the problem. The great discovery of RMT is that many observables are the same for various statistical ensembles of random matrices.

To illustrate this, let us cite two classical results of RMT. The eigenvalue density of a real symmetric or complex Hermitian $N\times N$ matrix, whose entries in the upper/lower triangle are independent, identically distributed random variables
with a finite variance equal to $\sigma^2/N$, converges for $N\rightarrow \infty$ to a limiting distribution
\bq
\rho(\lambda) = \frac{1}{2\pi\sigma^2}\sqrt{4\sigma^2-\lambda^2},  \quad
\mbox{for} \ \lambda \in [-2\sigma,2\sigma],
\label{wigner}
\eq
known as Wigner's semicircle distribution,
one of the best known results of classical RMT. The class
of matrices whose spectrum converges to the limit law (\ref{wigner}) 
is actually much broader and embraces matrices with entries being independent random variables which fulfill Pastur-Lindeberg's condition \cite{PASTUR}. 
This is an example of macroscopic universality of random matrices. In this
paper we concentrate on macroscopic properties and do not discuss microscopic properties of eigenvalue statistics.

An analogous formula for a non-Hermitian random matrix, which is another
example of a macroscopic law, reads 
\bq
\rho(z,\bar{z}) = \left\{ \begin{array}{cc} \frac{1}{\pi \sigma^2} \ 
& \mbox{for} \quad |z|\le \sigma,\\ & \\
0 \ & \mbox{for} \quad |z|>\sigma, \end{array} \right. \label{gg_distr}
\eq
where $z=x+iy$ is a complex number. 
The distribution (\ref{gg_distr}) is called Girko-Ginibre's distribution. The eigenvalue density 
has a rotational symmetry in the complex plane and is uniform inside the circle of radius $\sigma$. 
More generally, if a matrix has independent but not identically
distributed Hermitian and anti-Hermitian degrees of freedom \cite{SOMMERS}, the limit 
law (\ref{gg_distr}) assumes an elliptic form 
\bq
\rho(z,\bar{z}) = \left\{ \begin{array}{ll} \frac{1}{(1-\tau^2) \pi \sigma^2} \ 
& \mbox{for} \quad \frac{x^2}{\sigma^2(1+\tau)^2} + \frac{y^2}{\sigma^2(1-\tau)^2}  
\le 1, \\ & \\ 0 \ & \mbox{otherwise}, \end{array} \right. \label{elliptic_distr}
\eq
where $\sigma^2>0$ is an effective scale parameter  
and $\tau \in [-1,1]$ is a flatness of the ellipse. 
For $\tau=0$ one recovers the circular law 
(\ref{gg_distr}). For $\tau \rightarrow \pm 1$ the support of the distribution
(\ref{elliptic_distr}) reduces to a cut $[-2\sigma,2\sigma]$ 
on the real (for $\tau \rightarrow 1$) or imaginary (for $\tau \rightarrow - 1$) axis and the distribution itself reduces to 
a Wigner law (\ref{wigner}), as one can see by projecting
the elliptic distribution (\ref{elliptic_distr}) onto the real 
(imaginary) axis before taking the limit $\tau\rightarrow \pm 1$.

It might be striking that the derivation of the (apparently simple) functional form of $\rho(z,\bar{z})$ 
for the Girko-Ginibre ensemble is less straightforward than the one for the (more complex) Wigner semicircle law. The reason is that there are many powerful methods invented for Hermitian random matrices: via orthogonal polynomials or Selberg's integral \cite{mehta}, supersymmetric method 
\cite{efetov}, diagrammatic expansion \cite{feinberg}, Dyson gas \cite{dyson} and free random variables \cite{frv}.

In this paper we would like to present a result for non-Hermitian random matrices which is to a large extent universal, similarly to the two classical examples cited above. We shall show that the eigenvalue density $\rho_\X(z,\bar{z})$ of a product 
\bq
\X=\X_1 \X_2 \cdots \X_M,
\label{product}
\eq 
of $M\ge 2$ independent $N\times N$ Gaussian matrices for which $\left<\X_{1,ij}\right>=\dots=\left<\X_{M,ij}\right>=0$ and $\left<|\X_{1,ij}|^2\right>=\sigma_1^2/N,\dots,\left<|\X_{M,ij}|^2\right>=\sigma_M^2/N$ for all $i,j$, assumes in the limit of $N\rightarrow \infty$ the following form:
\bq
\rho_\X(z,\bar{z}) = 
\left\{ 
\begin{array}{ll} 
\frac{1}{M\pi} \sigma^{-\frac{2}{M}} |z|^{-2+\frac{2}{M}} & \mbox{for} \ |z|\le \sigma,
\\ & \\ 0 & \mbox{for} \ |z|>\sigma,
\end{array} \right.
\label{main}
\eq
where the effective scale parameter $\sigma = \sigma_1 \sigma_2 \ldots \sigma_M$. 
This surprisingly simple formula
is the main result of our paper. What is even more surprising is that this 
formula holds for a product of independent but not identically distributed
Gaussian matrices. This means that the individual matrices $\X_i$'s in the product
may come from different Gaussian ensembles (GUE, GOE or various elliptic Gaussian 
non-Hermitian matrices) and the eigenvalue density will always be given
by (\ref{main}). In other words, 
even if $\X_1,\dots,\X_M$ have oblate eigenvalue spectra, with
$\tau_1\ne 0$, $\ldots$ , $\tau_M\ne 0$, their product will have a rotationally-symmetric one.
We shall derivate this result with help of a diagrammatic technique appropriately tailored to non-Hermitian random matrices 
\cite{NONHER,DIAG} and to products of random matrices \cite{rjanik}. 
In order to make the paper self-contained we will also give an introduction to the 
diagrammatic methods (for a brief review see also \cite{review1}). 

It is tempting to conjecture that the limit law for the product
(\ref{main}) holds also for a wider class of matrices, including
Wigner matrices whose elements are independent, 
identically distributed random variables with a finite variance
or more generally, for matrices which fulfill the Pastur-Lindeberg's
condition \cite{PASTUR}. We will present a numerical support for this
conjecture.

The second objective of this paper is to use (\ref{main}) in order to verify an interesting 
conjecture made in Ref.~\cite{biely} saying that if the eigenvalue density $\rho(x,y)$ of 
a non-Hermitian matrix $\X$ is rotationally symmetric on the complex plane $z=x+iy$, 
then the marginal distribution $\rho_*(x) = \int dy \rho(x,y)$ obtained by its projection
onto the real axis or a projection $\rho_*(y) = \int dx \rho(x,y)$ onto the imaginary axis 
must be equal to the eigenvalue density of the matrix $(\X+\X^\dagger)/\sqrt{8}$ or 
$i(\X-\X^\dagger)/\sqrt{8}$, respectively, both being Hermitian matrices. If true, this
would allow one to calculate $\rho(x,y)$ from $\rho_*(x)$ via the inverse Abel transform.
In particular, if one projects the Girko-Ginibre distribution (\ref{gg_distr}) onto the 
real (or imaginary) axis, one indeed obtains the Wigner semicircle law: 
$\rho_*(x) = \frac{2}{\pi\sigma^2}\sqrt{\sigma^2-x^2}$, which is the same as the 
eigenvalue density of the matrix $(\X+\X^\dagger)/\sqrt{8}$ (or $i(\X-\X^\dagger)/\sqrt{8}$). 
In \cite{biely} it was checked numerically that the relation seemed to apply also 
to more complicated ensembles. Here we shall present a counterexample by showing
that the projection of the eigenvalue density of a product $AB$ of two Hermitian matrices $A$ and
$B$ which is rotationally symmetric (\ref{main}) is different from the eigenvalue density of
the rescaled anti-commutator $(AB+BA)/\sqrt{8}$ and the commutator $i(AB-BA)/\sqrt{8}$, 
so the conjecture is not true.

\section{Generalities} 

\subsection{Eigenvalue density and the measure}
We are interested in the eigenvalue distribution of a random matrix $\X$ (\ref{product}) being a product of $M$ independent $N\times N$ real or complex Gaussian matrices.
The eigenvalues $\left\{\lambda_i\right\}$ of $\X$ are complex
since $\X$ may be general be non-Hermitian. The eigenvalue distribution is defined by
\bq
\rho_\X(z,\bar{z}) = 
\left\langle \frac{1}{N} \sum_{i=1}^N \delta^{(2)}(z-\lambda_i) \right\rangle,
\label{rho}
\eq
where $\bar{z}$ denotes complex conjugate of $z$. The averaging 
$\langle \ldots \rangle = \int \ldots d\mu (\X_1,\ldots, \X_M) $ 
is done with a factorized probability measure, which in the simplest
case of identically distributed matrices takes the form
\bq
d \mu(\X_1,\ldots,\X_M) \propto
\prod_{\mu=1}^M e^{-\frac{N\alpha}4 {\rm Tr} \X_\mu \X_\mu^\dagger} D\X_\mu,
\label{average}
\eq
where $D \X_\mu$ denotes a flat measure. This formula  applies to four generic
cases of $\X_\mu$ being (a) complex, (b) complex Hermitian, (c) real and
(d) real symmetric matrices. The parameter $\alpha$ is defined as 
$\alpha = \lim_{N\rightarrow \infty} 2 N_{dof}/N^2$ where $N_{dof}$
is the number of real degrees of freedom of the matrix $X$. For (a) the flat measure is given by
$D \X_\mu = \prod_{ij} {\rm d} X_{\mu,ij} {\rm d} \bar{X}_{\mu,ij}$ or equivalently by 
$D \X_\mu = \prod_{ij} {\rm d} ({\rm Re} X_{\mu,ij}) 
{\rm d} ({\rm Im} X_{\mu,ij})$ and $\alpha=4$; for (b) 
$D \X_\mu = \prod_i {\rm d} X_{ii} \prod_{i>j} {\rm d}({\rm Re} X_{\mu,ij}) 
{\rm d} ({\rm Im} X_{\mu,ij})$, $\alpha=2$; for (c) $D \X_\mu = \prod_{ij} {\rm d} X_{\mu,ij}$, $\alpha=2$ and finally for (d) $D \X_\mu = \prod_{i\ge j} {\rm d} X_{\mu,ij}$, $\alpha=1$. For (c) and (d) the Hermitian conjugate $\X_\mu^\dagger$ 
reduces to the transpose $\X_\mu^T$. The proportionality symbol in (\ref{average}) means that the measure is displayed without a normalization constant which is fixed by the condition $\int d \mu(\X_1,\ldots,\X_M) = 1$.

With this choice of $\alpha$ the variance of individual elements $\left<|\X_{\mu,ij}|^2\right>=1/N$ so that the scaling parameters $\sigma_1=\dots=\sigma_M=1$ and hence $\sigma=1$ in Eq.~(\ref{main}). This means that the eigenvalue density of individual matrices $\X_\mu$ is given by the Girko-Ginibre law (\ref{gg_distr}) for (a) and (c) and the Wigner law (\ref{wigner}) for (b) and (d), in both cases with $\sigma=1$. 
For sake of simplicity we stick to this choice in the rest of the paper. The spectrum for arbitrary $\sigma_1,\dots,\sigma_M$ can be obtained by a trivial rescaling.

Later on we will also consider a general case of matrices from the elliptic ensemble with the eigenvalue distribution (\ref{elliptic_distr}). 
We will also consider a product of non-identically distributed matrices,
where $\X_1,\dots,\X_M$ belong to different elliptic ensembles.

\subsection{The Green's function}
We shall follow here the standard strategy of calculating the eigenvalue density 
of a random matrix by first calculating the Green's function $g(z,\bar{z})$ and
then using an exact relation between the eigenvalue density and the Green's function.
Let us recall this relation. Using the following representation of the two-dimensional
delta function 
\bq 
\delta^{(2)}(z-\lambda) = \lim_{\epsilon \rightarrow 0} 
\frac{1}{\pi} \frac{\epsilon^2}{(|z-\lambda|^2 + \epsilon^2)^2} = \lim_{\epsilon \rightarrow 0} 
\frac{1}{\pi} \frac{\partial}{\partial \bar{z}} 
\left[\frac{\bar{z}-\bar{\lambda}}{|z-\lambda|^2 + \epsilon^2}\right],
\eq
one finds \cite{GINIBRE,GIRKO,SOMMERS,HAAKE} that
\bq
\rho_\X(z,\bar{z}) = \frac{1}{\pi} \frac{\partial g(z,\bar{z})}{\partial \bar{z}},
\label{rho_g}
\eq  
where 
\bq
g(z,\bar{z}) = \lim_{\epsilon\rightarrow 0} \left\langle
\frac{1}{N} \sum_i^N \frac{\bar{z}-\bar{\lambda}_i}{|z-\lambda_i|^2 + \epsilon^2} \right\rangle =
\lim_{\epsilon\rightarrow 0} \left\langle \frac{1}{N} {\rm Tr} 
\frac{\bar{z} \mathbbm{1}_N-X^\dagger}{(\bar{z}\mathbbm{1}_N-X^\dagger)(z\mathbbm{1}_N-X) + 
\epsilon^2 \mathbbm{1}_N} \right\rangle,
\label{Green}
\eq
and $\mathbbm{1}_N$ is an $N\times N$ identity matrix. As we shall see later,
the Green's function can be calculated in the limit $N\rightarrow \infty$ 
using a summation method for planar Feynman diagrams. It is convenient to think 
of $g(z,\bar{z})$ as a part of a larger object \cite{ZEENEW2}, a $2N\times 2N$ matrix $G$ with four $N\times N$ blocks \cite{NONHER,DIAG}:
\bq
G= \left( \begin{array}{cc} G_{zz} & G_{z\bar{z}} \\ 
G_{\bar{z}z}& G_{\bar{z}\bar{z}} \end{array} \right) =
\lim_{\epsilon \rightarrow 0} \left\langle 
\left( \begin{array}{cc} z \mathbbm{1}_N - X & i\epsilon \mathbbm{1}_N \\ 
i\epsilon \mathbbm{1}_N & \bar{z} \mathbbm{1}_N - X^\dagger \end{array} \right)^{-1}
\right\rangle.
\label{G}
\eq
Before we continue let us shortly comment on the notation used in the last formula,
since we will also use it in the remaining part of the paper. 
The subscripts $zz$, $z\bar{z}$, $\bar{z}z$ and $\bar{z}\bar{z}$ 
refer to the position of the $N\times N$ blocks in the corresponding $2N\times 2N$ 
matrix. In the shorthand notation the arguments $(z,\bar{z})$ of a function 
defined on the complex plane are skipped, so the correct reading of, for 
instance, $G_{zz}$ is $G_{zz}=G_{zz}(z,\bar{z})$. 
We will also use a convention that the normalized trace of an $N\times N$ matrix 
denoted by a capital letter will be denoted by the corresponding small 
letter, for instance $g_{z\bar{z}} = \frac{1}{N} {\rm Tr} G_{z\bar{z}}$.

Now coming back to the problem, by inverting 
the matrix in the brackets on the right-hand side in the last equation we can see
that the Green's function $g(z,\bar{z})$ is equal to the normalized trace
of the upper-left sub-matrix,
\bq
g(z,\bar{z}) \equiv g_{zz}(z,\bar{z}) = \frac{1}{N} {\rm Tr} \; G_{zz}(z,\bar{z}).
\eq 
When one calculates the Green's function (\ref{Green}) or the matrix $G$ (\ref{G}),
one has to take the limit $N\rightarrow \infty$ first, and only then allow for $\epsilon\to 0$. 
This comes from the following reasoning. If $\epsilon=0$, for finite $N$ the 
function in the brackets $\langle \ldots \rangle$ on the right hand side of (\ref{Green})
has isolated poles on the complex plane. However, in the limit $N\rightarrow \infty$ the poles 
coalesce and the function becomes non-holomorphic. 
One cannot then make an analytic continuation of the function from holomorphic to nonholomorphic region,
as it is done when calculating $G$ by diagrammatic method which utilizes $O(1/z)$ expansion.
A small $\epsilon>0$ is necessary to make $G$ analytic everywhere.
If one naively first took the limit $\epsilon \rightarrow 0$ and only then the limit 
$N\rightarrow \infty$, the matrix $G$ would become block-diagonal: 
$G_{zz} =  \langle (z-X)^{-1} \rangle$, 
$G^{\dagger}_{\bar{z}\bar{z}} = \langle (\bar{z}-X^\dagger)^{-1} \rangle$
and $G_{z\bar{z}}=G_{\bar{z}z} = 0$. However, we shall see that
\bq
g_{z\bar{z}}(z,\bar{z}) = 
\lim_{\epsilon\rightarrow 0} \lim_{N\rightarrow \infty} \left\langle \frac{1}{N} {\rm Tr} 
\frac{-i\epsilon \mathbbm{1}_N}{(\bar{z}\mathbbm{1}_N-X^\dagger)(z\mathbbm{1}_N-X) + 
\epsilon^2 \mathbbm{1}_N} \right\rangle
\label{Order}
\eq
and $g_{\bar{z}z}(z,\bar{z})$ differ from zero in the non-holomorphic region.
In Ref.~\cite{NONHER} it was shown that these quantities are related to the statistics of left and right eigenvectors of the non-Hermitian random matrix ensemble.

The quantities $g_{z\bar{z}}=g_{\bar{z}z}$ are purely imaginary,
and $\gamma= -g_{z\bar{z}}g_{\bar{z}z}$ is a sort of order 
parameter for non-holomorphic behavior, which is positive in a region of the complex plane 
where the Green's function is non-holomorphic. The effect of pole coalescence 
and the emergence of a non-holomorphic behavior is very similar to 
the spontaneous breaking of a global symmetry in statistical models.
In such systems the symmetry is preserved as long as the system size $N$
is finite. It may, however, get spontaneously broken in the limit 
$N\rightarrow \infty$. Let us take the Ising model as an example.
Its Hamiltonian is invariant under a global transformation flipping
all spins and hence it has a $Z_2$ symmetry. 
As long as the number of spins is finite, the system is $Z_2$-symmetric and the average magnetization, which is an order 
parameter, is equal zero. However in the thermodynamic limit, 
that is when the system size becomes infinite, 
the $Z_2$ symmetry gets spontaneously broken below a 
critical temperature and the average magnetization is non-zero. 
If one first calculated the average magnetization for a finite system 
and only then took the limit $N\rightarrow \infty$, the magnetization would
be zero in this limit for all temperatures. To avoid the problem one can 
introduce a tiny external magnetic field $h$ which weakly breaks the symmetry for 
finite-size systems. Now, if one first takes the limit $N \rightarrow \infty$ 
and only then $h \rightarrow 0$, one will obtain the correct result.
In our case, the small parameter $\epsilon$ plays an analogous role to $h$ and it guaranties 
that non-holomorphic contributions will be correctly picked up for $N\rightarrow \infty$.

\subsection{Linearization}
Let us have a closer look at the function in the brackets 
in the definition of the Green's function (\ref{Green}). 
In our original problem the matrix $X$ is a product $X = X_1\ldots X_M$
of random matrices so it is a non-linear object from the point 
of view of the degrees of freedom that one has to average over.
As a consequence the diagrammatic method 
would become very complicated. 
One can, however, linearize the problem by a trick used in \cite{rjanik} which relies on substituting $\X$ by a matrix $\Y$ of dimensions 
$MN\times MN$ which is linear in $X_k$'s and has eigenvalues 
closely related to those of $\X$. The matrix $\Y$ is constructed
from $X_\mu$'s which are placed in a cyclic positions of 
a sparse $MN\times MN$ matrix,
\bq
Y = \left( \begin{array}{lllllc}
                         0 &  X_1 &     &            & 0       \\
                         0 & 0    & X_2 &            & 0        \\      
                           &      & \ddots & \ddots     &         \\
                         0 &      &     &          0 & X_{M-1} \\
                       X_M &      &     &            &  0      
\end{array}\right).
\label{Y}
\eq
One can immediately discover a relation between eigenvalues of $\Y$ 
and those of $X=X_1\ldots X_M$ if one calculates the $M$-th power $\Y$
which gives a block-diagonal matrix
\bq
Y^M = \left( \begin{array}{cccc}
                          Y_1 &      &  &         0 \\
                              &  Y_2 &  &          \\      
                              &      &  \ddots  &         \\
                           0  &      &          & Y_M 
                  
\end{array}\right),
\eq
with $Y_\mu$ being cyclic permutations of $X_\mu$'s, 
$Y_\mu=X_\mu X_{\mu+1} \ldots X_{\mu+M-1}$ (in the cyclic convention $X_{\mu+M}\equiv X_\mu$,
and $X_0\equiv X_M$). It is easy to see that all blocks $Y_\mu$ have the 
same eigenvalues. Indeed, if $\lambda$ is an eigenvalue of $Y_\mu$ to an 
eigenvector $\vec{v}_\mu$, $Y_\mu\vec{v}_\mu=\lambda \vec{v}_\mu$, it is also an 
eigenvalue of $Y_{\mu-1}$ to the eigenvector $\vec{v}_{\mu-1} = X_{\mu-1} \vec{v}_\mu$.
One can see this by multiplying both sides $Y_\mu\vec{v}_\mu=\lambda \vec{v}_\mu$ by $X_{\mu-1}$, obtaining $X_{\mu-1} Y_\mu \vec{v}_\mu = \lambda X_{\mu-1} \vec{v}_\mu$ which is equivalent to $Y_{\mu-1} \vec{v}_{\mu-1} = \lambda \vec{v}_{\mu-1}$. 
In other words, the matrix $\Y^M$ has exactly the same eigenvalues 
as $\X$ and each eigenvalue is $M$-fold degenerated.
Eigenvalues of $\X$ are thus related to those of $\Y$ as $\lambda_X=\lambda_Y^M$.
The eigenvalue density $\rho_\X(z,\bar{z})$ can be calculated from 
$\rho_\Y(w,\bar{w})$ of $\Y$ by changing the variables $z=w^M$:
\bq
	\rho_\X(z,\bar{z}) = M \frac{\partial w}{\partial z} \frac{\partial \bar{w}}{\partial \bar{z}} \; \rho_\Y(w,\bar{w})
	=\frac{1}{M} |z|^{-2+\frac{2}{M}} \rho_\Y(w(z),\bar{w}(\bar{z})) \label{rho_X}.
\eq
The factor $M$ in front of the Jacobian is related to the fact that the transformation 
$z=w^M$ maps the complex plane $M$ times onto itself. 
The problem is thus reduced to finding the spectral density of $\Y$, which is linear with respect to $X_1,\dots,X_M$.
The density $\rho_\Y(w,\bar{w})$ can be found from the appropriate Green's function. We will show below that $\rho_\Y(w,\bar{w})$ is given by a Girko-Ginibre distribution (\ref{gg_distr}), irrespectively of $M$ and of  $\tau_1$, $\tau_2$ $\ldots$ and $\tau_M$. This is a general result. 
In particular, for $M=2$ the matrix $Y$ (\ref{Y}) has an anti-diagonal block structure as chiral Gaussian matrices which have been intensively studied in the context of spectral properties of the Dirac operator in QCD \cite{VZ}. In this case, the form of the eigenvalue density of $Y$ for circular case ($\tau_1=\tau_2=0$) can be inferred from  results presented in \cite{O,A,APS} for complex, quaternion real, and real  matrices, respectively.

\section{Green's function and planar diagrams}
In this section we recall the diagrammatic technique of calculating the Green's function.
We begin with Hermitian matrices and later generalize the 
method to non-Hermitian ones and eventually to matrices
which additionally have a block structure like the matrix $\Y$ 
from the previous section.

Let us make a general comment before we proceed.
The diagrammatic method is based on the observation that the Green's
function $G$ can be interpreted as a generating function for 
connected two-point Feynman diagrams. In the limit $N \rightarrow \infty$
only planar diagrams contribute to $G$ since 
non-planar ones are suppressed by at least a factor $O(1/N)$ \cite{BIPZ,THOOFT}. 
In this limit one can write a set of two self-consistent algebraic
matrix equations which relate $G$ to a generating function, $\Sigma$, for
one-line irreducible diagrams. The equations are shown
schematically in Fig. \ref{fig1} and will be explained later. They can be solved
for $G$. We want to stress that these equations have exactly the same 
form for Hermitian, complex matrices and for matrices with a block structure. 
They only differ by an algebraic structure reflecting indexing of the 
matrices $G$ and $\Sigma$.

We finish with a remark that these equations hold for $N\rightarrow \infty$.
In the context of the discussion about the order of taking the limits in 
(\ref{Order}) this means that one can safely set $\epsilon=0$ 
since the limit $N\rightarrow \infty$ has already been taken.  
    
\subsection{Hermitian matrices}
\begin{figure}
	\includegraphics[width=16cm, bb=53 242 562 814]{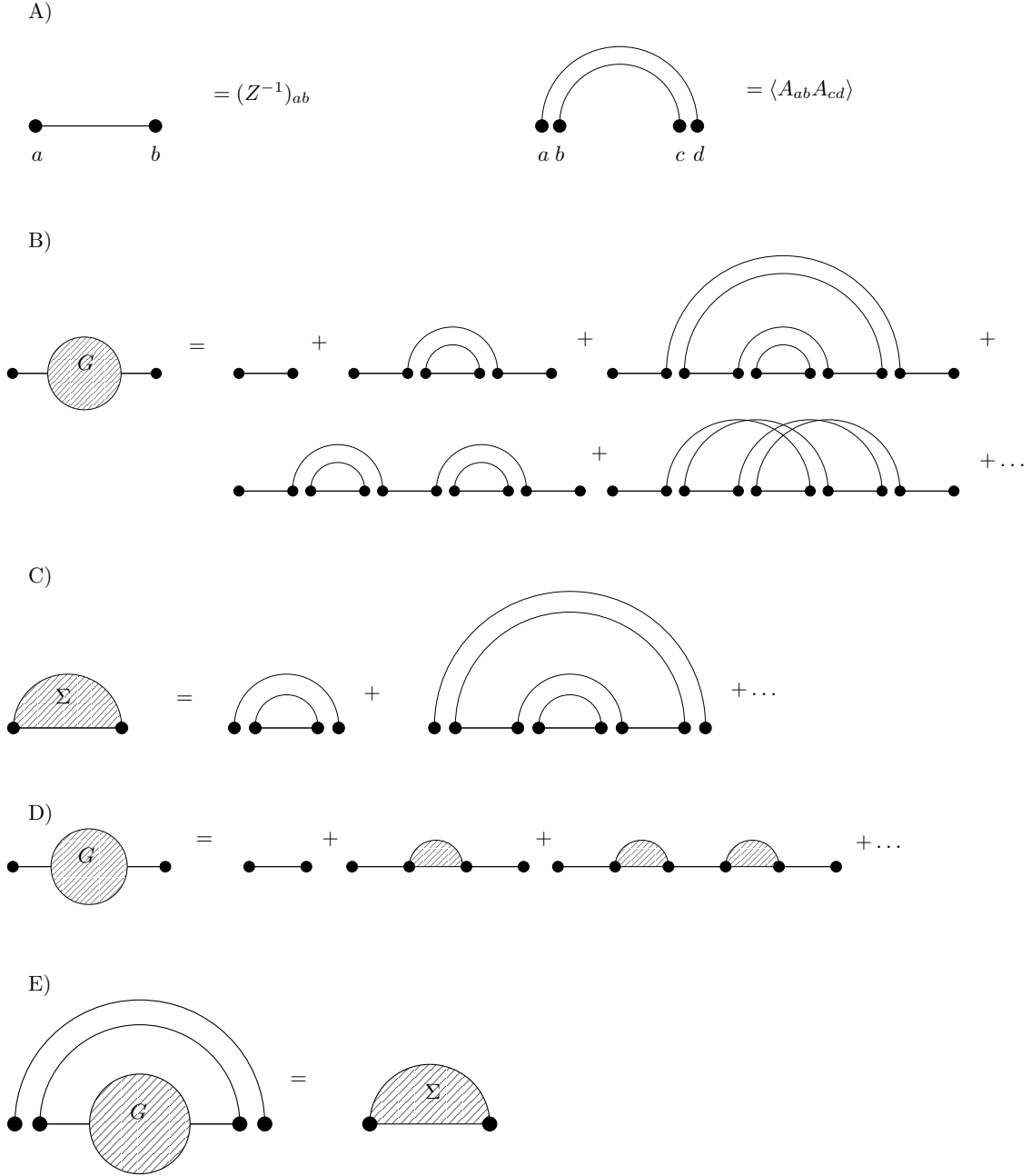}
	\caption{\label{fig1} (A) Feynman rules.
$(Z^{-1})_{ab}$ is drawn as a line between $a$ and $b$ and the propagator
$\langle A_{ab} A_{cd} \rangle$
as a double arc joining $a$ with $d$ and 
$b$ with $c$, respectively. (B) Graphical representation of
Eq.~(\ref{expand}). The last three displayed graphs correspond to the third
term in (\ref{expand}). The contribution of the last diagram can be
neglected in the large $N$ limit since it is non-planar and has a suppressing factor $1/N^2$.
(C) Definition of self-energy $\Sigma$. 
(D) The first Dyson-Schwinger equation which relates $G$ to $\Sigma$. 
(E) The second Dyson-Schwinger equation.
}
\end{figure}

We will first demonstrate the diagrammatic technique on the example of Hermitian matrices and derive the Wigner semicircle law (\ref{wigner}).
Let us assume that $A=A^\dagger$, $A=\{A_{ab}\},a=1,\ldots,N, b=1,\ldots,N$ is drawn from an ensemble with a probability measure
\bq
d\mu(A) \propto e^{-\frac{N}{2} {\rm Tr} A^2} DA,
\eq
where $DA = \prod_a {\rm d} A_{aa} \prod_{a>b} {\rm d} ({\rm Re} A_{ab}) {\rm d} ({\rm Im} A_{ab})$. 
The normalization constant, which is implicit in the above formula, 
is fixed by the condition $\int d\mu (A)=1$. 
The eigenvalues $\lambda_i$ of the matrix $A$
are real. This makes the situation simpler than the one for general non-Hermitian matrices discussed in Sec. II. The eigenvalue density can be expressed as \cite{guhr}
\bq
\rho(\lambda) = \left\langle \frac{1}{N} \sum_i^N \delta(\lambda-\lambda_i)\right\rangle,
\eq
where now the delta function is one-dimensional. Also
the Green's function $G$ matrix takes a simpler form,
\bq
G = \left\langle (Z-A)^{-1} \right\rangle \equiv \int (Z-A)^{-1} d\mu(A).
\label{Gz}
\eq
Here $Z= z \mathbbm{1}_N$, where $z$ is a complex number. The Green's function $g(z) \equiv \frac{1}{N} {\rm Tr}\, G(z)$ is obtained by the Stieltjes transform of the eigenvalue density:
\bq
 g(z) = \int d\lambda \frac{\rho(\lambda)}{z-\lambda}.
\eq
The last equation yields:
\bq
\rho(\lambda) = -\frac{1}{\pi} {\rm Im}\, g(\lambda + i\epsilon),
\label{rhog}
\eq
for $\epsilon\to 0$, as follows from a standard representation of the one-dimensional delta function
$\delta(x) = -\frac{1}{\pi} {\rm Im} (x + i\epsilon)^{-1}$. 
The above Green's function can be calculated analytically in the large $N$ limit, expanding (\ref{Gz}) in terms of powers of $Z^{-1}$:
\bq
G(z) = Z^{-1} + \langle Z^{-1} A \; Z^{-1} A \; Z^{-1} \rangle + 
\langle Z^{-1} A \; Z^{-1} A \; Z^{-1} A \; Z^{-1} A \; Z^{-1} \rangle + \ldots
\label{expand}
\eq
Factors $Z^{-1}$ are independent of $A$'s and thus can be pulled out
of the average brackets. What remains are correlation functions of the type $\langle A_{i_1 i_2} \ldots A_{i_{2n-1} i_{2n}} \rangle$ which 
by virtue of the Wick theorem can be expressed as products of two-point 
correlation  functions (propagators)
\bq
\langle A_{ab} A_{cd} \rangle = \frac{1}{N} \delta_{ad} \delta_{bc} \ .
\label{corrA}
\eq
This observation allows one to graphically represent equation (\ref{expand}) 
as a sum over Feynman diagrams (see for instance \cite{BGJJ}), as shown in Fig. \ref{fig1}B. Each propagator is represented as a double arc joining two pairs of matrix indices, while $Z^{-1}_{ab}$ is drawn as a horizontal line joining indices $a$ and $b$ (Fig. \ref{fig1}A).
In order to calculate $G_{ab}$ one has to sum up contributions of all connected 
diagrams with two external points $a, b$. For finite $N$ this is not an easy task
because there are infinitely many diagrams. The problem enormously 
simplifies in the limit $N\rightarrow \infty$ since in this limit only planar 
diagrams contribute to the leading term of $1/N$ expansion and all non-planar
diagrams can be neglected \cite{BIPZ,THOOFT}. It turns out that all planar diagrams can be summed 
up using an old trick known from field theory which reduces the problem 
to a closed set of equations for $G$. These equations are known as Dyson-Schwinger equations and we will discuss them now. 

First, we introduce a generating function $\Sigma$ for  one-line irreducible diagrams, that is diagrams which 
cannot be split by cutting a single horizontal line (see Fig.~\ref{fig1}C).
$\Sigma_{ab}$ generates all one-line irreducible diagrams with vertices $a$ and $b$. 
The two generating functions are related to each other because any diagram 
from $G$ can be constructed as a sandwich of horizontal lines and one-line 
irreducible diagrams (Fig.~\ref{fig1}D):
\bq
G = Z^{-1} + Z^{-1} \; \Sigma \; Z^{-1} + 
Z^{-1} \; \Sigma Z^{-1} \; \Sigma \; Z^{-1} \; \Sigma Z^{-1}
+ \ldots = \left(Z - \Sigma \right)^{-1}.
\label{ds1}
\eq
This matrix equation can be viewed as a definition of $\Sigma$.
The introduction of $\Sigma$ itself does not help to solve the problem.
However, one can write down an independent equation for $\Sigma$ and $G$. 
It follows from the observation that any one-line irreducible 
diagram can be obtained from a diagram from $G$ by adding an arc 
(a propagator) to it (Fig.~\ref{fig1}E). This gives
\bq
\Sigma_{ab}  = \sum_{c,d} G_{cd} \frac{1}{N} \delta_{cd} \delta_{ab} = g \delta_{ab},
\label{ds2}
\eq
or, in matrix notation, $\Sigma=g\mathbbm{1}_N$. Taking trace of both sides we
obtain $\sigma=g$ where $\sigma \equiv \frac{1}{N} {\rm Tr} \Sigma$ is the
normalized trace of $\Sigma$. The two equations (\ref{ds1}) and (\ref{ds2})
form a closed set of equations which can be solved for the Green's function $g(z)$.
Inserting the last equation to (\ref{ds1}) with $Z=z\mathbbm{1}_N $ we have 
$g\left(z-g\right)=1$ and hence $g(z) = \frac{1}{2}(z - \sqrt{z^2-4})$ and 
$\rho(\lambda) = \frac{1}{2\pi}\sqrt{4-\lambda^2}$, as follows from (\ref{rhog}). 
 
\subsection{Complex matrices}
Let us now discuss how to calculate the Green's function in case of non-Hermitian Gaussian random matrices
with complex entries (see for instance \cite{review1}). 
The probability measure is now
\bq
d\mu(\A)  \propto e^{- N {\rm Tr} \A\A^\dagger} \prod_{i,j} {\rm d}({\rm Re} A_{ij}) {\rm d}({\rm Im} A_{ij}),
\label{PXX}
\eq
which corresponds to $\alpha=4$ in Eq.~(\ref{average}). The propagators are
\bq
\begin{array}{lcl}
\langle A_{ab} A_{cd} \rangle = 0 & , & 
\langle A_{ab} A^\dagger_{cd} \rangle = 
\frac{1}{N} \delta_{ad} \delta_{bc}, \\ &  & \\
\langle A^\dagger_{ab} A_{cd} \rangle = 
\frac{1}{N} \delta_{ad} \delta_{bc} & , & 
\langle A^\dagger_{ab} A^\dagger_{cd} \rangle = 0. 
\end{array}
\label{p2x2}
\eq
It is convenient to think of $\A$ and $\A^\dagger$ as $N\times N$
sub-matrices of a $2N \times 2N$ matrix
\bq
{\cal A} = \left( \begin{array}{cc} {\cal A}_{zz} & {\cal A}_{z\bar{z}} \\ 
{\cal A}_{\bar{z}z}& {\cal A}_{\bar{z}\bar{z}} \end{array} \right) =
\left( \begin{array}{cc} \A & 0 \\ 
0 & \A^\dagger \end{array} \right).
\label{calX}
\eq
The off-diagonal blocks are equal zero for this particular matrix.  
We use a convention discussed in Section II: the position of 
an $N \times N$ sub-matrix is denoted
by subscripts $z,\bar{z}$.  We apply the same notation 
to other $2N \times 2N$ matrices: the Green's function, 
the self-energy $\Sigma$ and the matrix $Z$,
\bq
G= \left( \begin{array}{cc} G_{zz} & G_{z\bar{z}} \\ 
G_{\bar{z}z}& G_{\bar{z}\bar{z}} \end{array} \right), \qquad
\Sigma= \left( \begin{array}{cc} \Sigma_{zz} & \Sigma_{z\bar{z}} \\ 
\Sigma_{\bar{z}z}& \Sigma_{\bar{z}\bar{z}} \end{array} \right)
, \qquad 
Z= \left( \begin{array}{cc} Z_{zz} & Z_{z\bar{z}} \\ 
Z_{\bar{z}z}& Z_{\bar{z}\bar{z}} \end{array} \right).
\label{G2N}
\eq
Matrix elements of the block $G_{zz}$ of $G$ will be denoted by 
$G_{ab}$, elements of $G_{z\bar{z}}$ by $G_{a\bar{b}}$, etc.
In other words, the subscripts $z$ and $\bar{z}$ serve also as templates 
for the corresponding barred or unbarred indices. 
For completeness let us rewrite the propagators (\ref{p2x2}) using this notation:
\bq
\begin{array}{lcl}
\langle {\cal A}_{ab} {\cal A}_{cd} \rangle = 0 & , & 
\langle {\cal A}_{ab} {\cal A}_{\bar{c}\bar{d}} \rangle = 
\frac{1}{N} \delta_{a\bar{d}} \delta_{b\bar{c}}, \\ &  & \\
\langle {\cal A}_{\bar{a}\bar{b}} {\cal A}_{cd} \rangle = 
\frac{1}{N} \delta_{\bar{a}d} \delta_{\bar{b}c} & , & 
\langle {\cal A}_{\bar{a}\bar{b}} {\cal A}_{\bar{c}\bar{d}} \rangle = 0.
\end{array}
\label{XXprop}
\eq

Now we are ready to write down Dyson-Schwinger equations for complex matrices. 
The first equation is identical to Eq.~(\ref{ds1}), except that now 
$G$, $\Sigma$ and $Z$ have dimensions $2N\times 2N$:
\bq
\left( \begin{array}{cc} G_{zz} & G_{z\bar{z}} \\ 
G_{\bar{z}z}& G_{\bar{z}\bar{z}} \end{array} \right) =
\left( \begin{array}{cc} Z_{zz} \!-\! \Sigma_{zz} & Z_{z\bar{z}} \!-\! \Sigma_{z\bar{z}} \\ 
Z_{\bar{z}z} \!-\! \Sigma_{\bar{z}z}& Z_{\bar{z}\bar{z}} \!-\! \Sigma_{\bar{z}\bar{z}} 
\end{array} \right)^{-1}.
\label{DS1}
\eq
This equation is general, but later we will write it for a specific form of $Z$ 
relevant for the calculation of the eigenvalue density. The second equation, which 
corresponds to (\ref{ds2}), can be derived using the propagators defined in Eq.~(\ref{XXprop}).
It can be done separately in each sector $zz$, $z\bar{z}$, $\bar{z}z$ and $\bar{z}\bar{z}$:
\bq
\begin{array}{lcl}
\Sigma_{ad} = 0 & , & 
\Sigma_{a\bar{d}} = \frac{1}{N} \delta_{a\bar{d}} \delta_{b\bar{c}} G_{b\bar{c}} = 
\delta_{a\bar{d}} g_{z\bar{z}}, \\ &  & \\
\Sigma_{\bar{a}d} = \frac{1}{N} \delta_{\bar{a}d} \delta_{\bar{b}c} G_{\bar{b}c} = 
\delta_{\bar{a}d} g_{\bar{z}z}
& , & 
\Sigma_{\bar{a}\bar{d}} = 0,
\end{array}
\eq 
where $g_{z\bar{z}}=\frac{1}{N} {\rm Tr} G_{z\bar{z}}$ and
$g_{\bar{z}z}=\frac{1}{N} {\rm Tr} G_{\bar{z}z}$.
In matrix notation the last equation can be written as
\bq
\left( \begin{array}{cc} \Sigma_{zz} & \Sigma_{z\bar{z}} \\ 
\Sigma_{\bar{z}z}& \Sigma_{\bar{z}\bar{z}} \end{array} \right) =
\left( \begin{array}{cc} 0  & g_{z\bar{z}} \mathbbm{1}_N \\ 
g_{\bar{z}z} \mathbbm{1}_N & 0 \end{array} \right).
\label{DS2}
\eq
One should note that the form of this equation is independent of $Z$
while the form of the first Dyson-Schwinger equation (\ref{DS1}) is independent 
of the propagator structure. If we insert now 
\bq
	Z=\lim_{\epsilon\to 0} \left( \begin{array}{cc} z \mathbbm{1}_N & i\epsilon \mathbbm{1}_N \\ 
i\epsilon \mathbbm{1}_N & \bar{z} \mathbbm{1}_N \end{array} \right) \
= \left( \begin{array}{cc} z \mathbbm{1}_N  & 0 \\ 
0 & \bar{z} \mathbbm{1}_N  \end{array} \right)
\eq
to Eq.~(\ref{DS1}), remembering that we are allowed to take $\epsilon\to 0$
since all above equations are derived for large $N$ and hence the limit $N\to\infty$ has been
taken, we eventually obtain a matrix equation
\bq 
\left( \begin{array}{cc} G_{zz} & G_{z\bar{z}} \\ 
G_{\bar{z}z} & G_{\bar{z}\bar{z}} \end{array} \right) 
 = \left(\begin{array}{cc} z \mathbbm{1}_N - \Sigma_{zz} & -\Sigma_{z\bar{z}} \\ 
-\Sigma_{\bar{z}z} & \bar{z} \mathbbm{1}_N - 
\Sigma_{\bar{z}\bar{z}} \end{array} \right)^{-1},
\label{DS1final}
\eq
which together with (\ref{DS2}) forms a closed set of algebraic
equations for $G(z,\bar{z})$. 

We will now solve this set of equations and then determine $\rho(z,\bar{z})$ using Eq.~(\ref{rho_g}).
We first notice that  Eq.~(\ref{DS2}) reduces to 
a $2 \times 2$ matrix equation:
\bq
\left( \begin{array}{cc} \sigma_{zz} & \sigma_{z\bar{z}} \\ 
\sigma_{\bar{z}z}& \sigma_{\bar{z}\bar{z}} \end{array} \right) =
\left( \begin{array}{cc} 0  & g_{z\bar{z}} \\ 
g_{\bar{z}z} & 0 \end{array} \right),
\label{DSr1}
\eq
where, as before, small letters denote the normalized traces of the corresponding
blocks, for instance $\sigma_{zz} = \frac{1}{N} {\rm Tr} \Sigma_{zz}$.
Similarly, equation (\ref{DS1final}) reduces to 
\bq
\left( \begin{array}{cc} g_{zz} & g_{z\bar{z}} \\ 
g_{\bar{z}z} & g_{\bar{z}\bar{z}} \end{array} \right) 
 = \left(\begin{array}{cc} z  - \sigma_{zz} & -\sigma_{z\bar{z}} \\ 
-\sigma_{\bar{z}z} & \bar{z}  - \sigma_{\bar{z}\bar{z}} \end{array} \right)^{-1},
\label{DSr2}
\eq
which, after eliminating $\sigma$'s with help of Eq.~(\ref{DSr1}), leads to
\bq
\left( \begin{array}{cc} g_{zz} & g_{z\bar{z}} \\ 
g_{\bar{z}z} & g_{\bar{z}\bar{z}} \end{array} \right) =
\left(\begin{array}{cc} z & -g_{z\bar{z}} \\ 
-g_{\bar{z}z} & \bar{z} \end{array} \right)^{-1}
= \frac{1}{|z |^2 - g_{z\bar{z}} g_{\bar{z}z}} 
\left(\begin{array}{cc} \bar{z}   & g_{z\bar{z}} \\ g_{\bar{z}z} & z \end{array} \right).
\label{ggg}
\eq
This equation has two solutions. The first one corresponds to $g_{\bar{z}z}=g_{z\bar{z}}=0$ 
which gives $g_{zz} = z^{-1}$ and is equivalent to the trivial holomorphic solution and hence must be true for large $|z|$. The second solution corresponds to $|z|^2 - g_{z\bar{z}} g_{\bar{z}z}=1$.
In this case the off-diagonal blocks are different from zero and $g_{zz}=\bar{z}$. The two solutions match for $|z|^2=1$. 
Therefore, the first solution holds outside the unit circle and the second one inside the circle. 
Using the Gauss law (\ref{rho_g}) one finds
\bq
\rho(z,\bar{z}) = \left\{ 
\begin{array}{ll} \frac{1}{\pi}  & \mbox{for} \ |z|\le 1, \\ & \\ 0 & \mbox{for} \ |z|>1, 
\end{array}
\right.
\eq
which is the celebrated Girko-Ginibre distribution \cite{GINIBRE,GIRKO}. 

To summarize this part, one can write the closed set of algebraic
equations for $G$ and $\Sigma$
in the large-$N$ limit using diagrammatic relations between the generating 
function for connected two-point planar diagrams (given by $G$)
and the generating function for one-line irreducible two-point planar diagrams
(given by the free energy $\Sigma$). One can set $\epsilon=0$ in these
equations since they are derived already in the limit $N\rightarrow \infty$. 

\subsection{Complex matrices with a block structure}

\begin{figure}
	\includegraphics*[width=15.5cm]{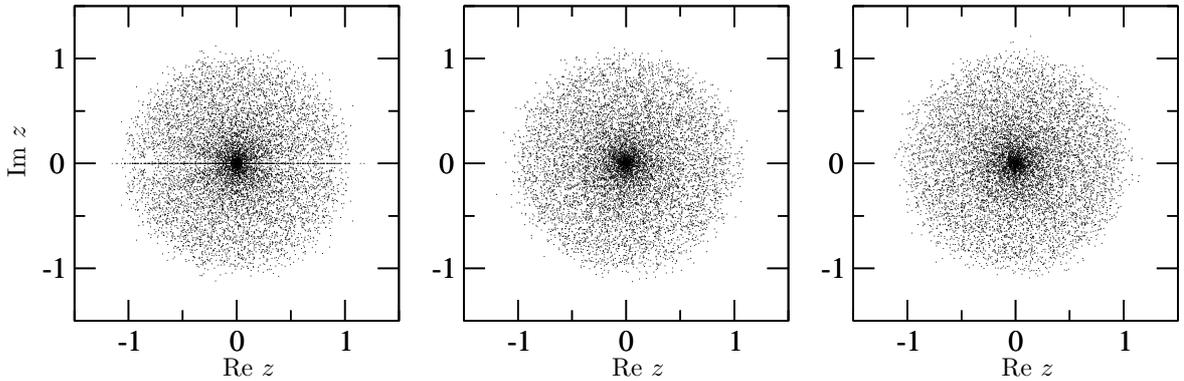}
	\caption{\label{fig2}Plots of $\rho_\X(z,\bar{z})$ for $\X_1,\X_2$ being two Hermitian matrices (left), two complex matrices (middle), and for $\X_1$ being a Hermitian and $\X_2$ an elliptic random matrix with $\phi=\pi/3$ (right). For each case 100 matrices of size $N=100$ were generated.}
\end{figure}

\begin{figure}
	\includegraphics*[width=15.5cm]{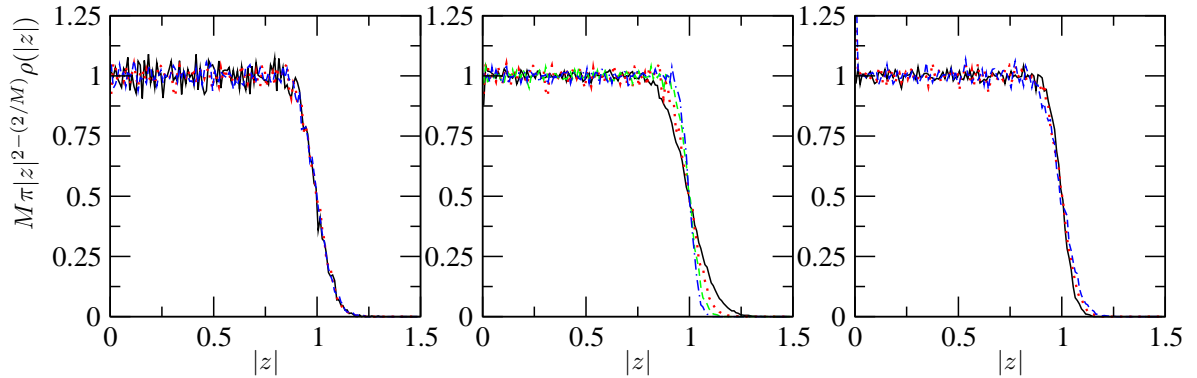}
	\caption{\label{fig3}Plots of $M\pi |z|^{2-\frac{2}{M}}\rho_\X(|z|)$ obtained from simulations for various $M$ and matrix sizes $N$. The theoretical distribution (not shown in the figure) which corresponds to (\ref{main}) is a step function $f(|z|)=1$ for $0<|z|<1$ and zero otherwise. Left: $\X=\X_1\X_2$ ($M=2$) for $N=100$ and $X_1,X_2$ taken from the same ensembles as in Fig. \ref{fig2}: black solid line for Hermitian, red dotted line for complex, and blue dashed line for Hermitian elliptic matrices. Middle: $M=2$, complex matrices of size $N=50,100,200,400$ (black solid, red dotted, green dashed and blue dotted-dashed lines, respectively). To obtain these plots, we averaged spectra of $10000,1000,1000$ and $500$ matrices and constructed histograms of absolute values of their eigenvalues. Right: $N=200$ and $M=2,3,4$ (black solid, red dotted and blue dashed lines). For each $M$, $1000$ matrices were generated.
	}
\end{figure}

We are now ready to calculate the Green's function $g_\Y(w,\bar{w})$ for the matrix 
$\Y$ (\ref{Y}) which has blocks $X_\mu$ being independent complex non-Hermitian Gaussian matrices \cite{rjanik}. 
The matrix $G$ will be now a $2NM\times 2NM$ matrix having four $NM\times NM$ blocks 
$G_{ww}$, $G_{w\bar{w}}$, $G_{\bar{w}w}$ and $G_{\bar{w}\bar{w}}$
which themselves consists of $M^2$ blocks of size $N\times N$ which we shall denote by
$G_{\mu\nu}$, $G_{\mu\bar{\nu}}$, $G_{\bar{\mu}\nu}$ and $G_{\bar{\mu}\bar{\nu}}$ 
respectively, for instance
\bq
G_{w\bar{w}} = \left( \begin{array}{lcl} 
G_{1\bar{1}} & \ldots & G_{1\bar{M}} \\
             & \ldots &              \\
G_{M\bar{1}} & \ldots & G_{M\bar{M}} \end{array} \right). 
\eq
There is an analogous block structure for the matrix $\Sigma$. 
One should distinguish Greek subscripts from Latin subscripts giving the position of the
matrix elements within the block. 
For instance, $\Sigma_{\mu\bar{\nu}}$
is an $N\times N$ sub-matrix of the block $\Sigma_{w\bar{w}}$ and 
$\left(\Sigma_{\mu\bar{\nu}}\right)_{a\bar{b}}$ is an element of this sub-matrix.
In this convention the normalized trace of a block
is $\sigma_{\mu\bar{\nu}} = \frac{1}{N} {\rm Tr} \Sigma_{\mu\bar{\nu}} = 
\frac{1}{N}\sum_{a=1}^N \left(\Sigma_{\mu\bar{\nu}}\right)_{a\bar{a}}$.
One can now repeat the same procedure which we applied to the matrix
having a single block and derive exact relations between the generating 
function $G$ and $\Sigma$ in the planar limit.  
The first Dyson-Schwinger equation,
\bq
\left( \begin{array}{cc} G_{ww} & G_{w\bar{w}} \\ 
G_{\bar{w}w} & G_{\bar{w}\bar{w}} \end{array} \right) 
 = \left(\begin{array}{cc} w \mathbbm{1}_{NM} - \Sigma_{ww} & -\Sigma_{w\bar{w}} \\ 
-\Sigma_{\bar{w}w} & \bar{w} \mathbbm{1}_{NM} - 
\Sigma_{\bar{w}\bar{w}} \end{array} \right)^{-1},
\label{Gww}
\eq
is almost identical as (\ref{DS1final}), except that the blocks and the
identity matrices are now of dimensions $NM\times NM$. To write the second equation,
we need to know the propagators. Let us first define a $2NM\times 2NM$ matrix ${\cal Y}$,
a counterpart of ${\cal A}$ from Eq.~(\ref{calX}):
\bq
{\cal Y} = \left( \begin{array}{cc} {\cal Y}_{ww} & {\cal Y}_{w\bar{w}} \\ 
{\cal Y}_{\bar{w}w}& {\cal Y}_{\bar{w}\bar{w}} \end{array} \right) =
\left( \begin{array}{cc} \Y & 0 \\ 
0 & \Y^\dagger \end{array} \right),
\label{calY}
\eq
where $\Y$ is cyclic as defined in Eq.~(\ref{Y}) and $\Y^\dagger$ is anti-cyclic,
\bq
\Y^\dagger = \left( \begin{array}{llllll}
                           0 &   &     &            & X_M^\dagger      \\
                           X_1^\dagger \quad  & 0      &         &   &   0     \\      
                           &           \ddots &   &   &         \\
                            &           &  X_{M-2}^\dagger       & 0 & \\
                           0 &           &         & X_{M-1}^\dagger  &  0      
\end{array}\right).
\label{Y2}
\eq
Since the block matrices ${\cal Y}_{\mu\mu+1}=X_\mu$ are assumed to be independent 
of each other,  the only non-zero propagators are
\bq
\langle {\cal Y}_{12,ab} {\cal Y}_{\bar{2}\bar{1},\bar{c}\bar{d}} \rangle = 
\langle {\cal Y}_{23,ab} {\cal Y}_{\bar{3}\bar{2},\bar{c}\bar{d}} \rangle = \ldots =
\langle {\cal Y}_{M1,ab} {\cal Y}_{\bar{1}\bar{M},\bar{c}\bar{d}} \rangle = \frac{1}{N} 
\delta_{a\bar{d}} \delta_{b\bar{c}},
\label{YYprop}
\eq
or in short 
\bq
\langle {\cal Y}_{12} {\cal Y}_{\bar{2}\bar{1}} \rangle = 
\langle {\cal Y}_{23} {\cal Y}_{\bar{3}\bar{2}} \rangle = \ldots =
\langle {\cal Y}_{M1} {\cal Y}_{\bar{1}\bar{M}} \rangle = \mathbbm{T},
\label{YYshort}
\eq
where the tensor $\mathbbm{T}$ has elements 
$T_{abcd} = \frac{1}{N}\delta_{ab} \delta_{cd}$, with indices corresponding
to the those of the matrices on the left-hand side. If we now insert these 
propagators to the second Dyson-Schwinger equation, we obtain
\bq
\Sigma_{\mu\bar{\mu}} = g_{\mu+1\overline{\mu+1}} \mathbbm{1}_N,
\label{Sg}
\eq
and $\Sigma_{\mu\bar{\nu}}=\Sigma_{\bar{\mu}\nu}=0$ for $\mu\neq\nu$.
The problem is symmetric with respect to permutation of the matrices
$X_\mu$, so  $g_{1\bar{1}}=\ldots=g_{M\bar{M}}\equiv g_{w\bar{w}}$ 
in the whole $w\bar{w}$-block and similarly in the $\bar{w}w$-block. 
Thus the last equation can be compactly written as
\bq
\Sigma_{w\bar{w}} = g_{w\bar{w}} \mathbbm{1}_{NM}, \qquad 
\Sigma_{\bar{w}w} = g_{\bar{w}w} \mathbbm{1}_{NM},
\label{Soffd}
\eq
where $\mathbbm{1}_{NM}$ is now the identity $NM\times NM$ matrix for the whole block,
$g_{w\bar{w}} = \frac{1}{NM} {\rm Tr} G_{w\bar{w}}$ and  
$g_{\bar{w}w} = \frac{1}{NM} {\rm Tr} G_{\bar{w}w}$. 
Inserting $\Sigma_{ww}=\Sigma_{\bar{w}\bar{w}}=0$ and (\ref{Soffd}) to (\ref{Gww}) we see that
each block on the right-hand side of (\ref{Gww}) is proportional to the identity 
matrix. Thus equation (\ref{Gww}) reduces to a $2\times 2$ matrix equation for the normalized
traces which play the role of proportionality coefficients at the identity matrices,
\bq
\left( \begin{array}{cc} g_{ww} & g_{w\bar{w}} \\ 
g_{\bar{w}w} & g_{\bar{w}\bar{w}} \end{array} \right) 
 = \left(\begin{array}{cc} w & -g_{w\bar{w}} \\ 
-g_{\bar{w}w} & \bar{w} \end{array} \right)^{-1}.
\eq
This is identical to (\ref{ggg}) for a complex matrix with a single
block discussed in the previous section. In other words,
the Green's function and hence also the eigenvalue density of the matrix $\Y$ does not 
depend on the number of blocks in $Y$ and is given by the Girko-Ginibre law \cite{GINIBRE,GIRKO}
\bq
\rho_\Y(w,\bar{w}) = \left\{ 
\begin{array}{ll} \frac{1}{\pi}  & \mbox{for} \ |w|\le 1, \\ & \\ 0 & \mbox{for} \ |w|>1. 
\end{array}
\right.
\eq
 This result is valid also for other matrices considered in Eq.~(\ref{average}), that is for real non-symmetric and Hermitian complex matrices, as long as $M>1$. It is so because what matters is the structure of propagators only, which is the same for all mentioned ensembles. 
 In particular, for $M=2$ one can deduce this formula from considerations of chiral ensembles \cite{O,A,APS}. In the next section we shall show how to derive the above result for the product of $M$ elliptic complex and/or real matrices with different oblateness parameters $\tau_1 \ne \ldots \ne \tau_M$. Now we will only observe that by inserting the Girko-Ginibre spectrum into Eq.~(\ref{rho_X}) we finally obtain
\bq
	\rho_X(z,\bar{z}) = \rho_X(|z|) = \left\{ \begin{array}{ll}
	\frac{1}{M\pi} |z|^{-2+\frac{2}{M}} & \mbox{for} \ |z|\le 1,
	\\ & \\ 0 & \mbox{for} \ |z|>1,
	\end{array}
	\right.  
\eq 
which completes the derivation of our main result.
In Figs.~\ref{fig2} and \ref{fig3} we show a comparison between the above formula and the spectrum of $\X$ obtained numerically by diagonalization of finite matrices. The agreement is very good. For the spectrum of the product of two Hermitian matrices (GUE) shown in the left panel of Fig.~\ref{fig2} we observe a small deviation from rotational symmetry manifesting as an accumulation of eigenvalues along the real axis and a depletion of eigenvalues in a narrow strip close to this axis. The number of eigenvalues on the
axis grows as $\sqrt{N}$ and the width of the strip decreases
as $1/\sqrt{N}$ when $N\rightarrow \infty$. This effect is almost identical as the one known for real Girko-Ginibre matrices \cite{E,AK}. If one multiplies three or more GUE matrices the effect disappears. A difference between the product of two and the product of more than two GUE matrices is that for two the trace ${\rm Tr} X_1 X_2$ is real whereas for three (or more) it is not.  In other words, the constraint of the trace to be real introduces a weak spherical symmetry breaking of the eigenvalue spectrum.

\section{Product of arbitrary Gaussian matrices (elliptic ensembles)}


Let us now consider a general class of non-Hermitian random matrices which include as special cases the well known examples of Hermitian (GUE), Girko-Ginibre, and anti-Hermitian ensembles. These ``elliptic'' ensembles were first introduced in \cite{SOMMERS} and can be defined as follows. A complex, elliptic matrix $\X$ is obtained as a linear
combination of two identical, independent Hermitian Gaussian matrices $A,B$:
$X = \cos(\phi) A + i \sin(\phi) B$, mixed with an arbitrary real mixing parameter $\phi$. 
Since $A$ and $B$ are independent, the corresponding propagators are 
$\langle A_{ab} A_{cd} \rangle = \frac{1}{N} \delta_{ad} \delta_{bc}$,
$\langle B_{ab} B_{cd} \rangle = \frac{1}{N} \delta_{ad} \delta_{bc}$, and
$\langle A_{ab} B_{cd} \rangle = 0$. When one changes variables from
$A$ and $B$ to $X$ and $X^\dagger$ one finds
\bq
\langle X_{ab} X_{cd} \rangle = \langle X^\dagger_{ab} X^\dagger_{cd} \rangle =
\tau \cdot \frac{1}{N} \delta_{ad} \delta_{bc} \ , \qquad
\langle X_{ab} X^\dagger_{cd} \rangle=\langle X^\dagger_{ab} X_{cd} \rangle = 
\frac{1}{N} \delta_{ad} \delta_{bc},
\label{tau_prop}
\eq
where $\tau = \cos(2\phi)$. The corresponding integration measure for $\X$ reads:
\bq
d\mu(\X) \propto \exp \left\{ - N \frac{1}{1-\tau^2} \left( 
{\rm Tr} XX^\dagger - \tau \frac{1}{2} {\rm Tr} \left(XX +X^\dagger X^\dagger\right)\right)\right\} \prod_{ij} {\rm d} ({\rm Re} X_{ij}) 
{\rm d} ({\rm Im} X_{ij}).
\label{Ptau}
\eq
For $\phi=0$ ($\tau=1$) the matrix $X$ is  Hermitian, for $\phi=\pi/2$ 
($\tau=-1$) it is anti-Hermitian while for $\phi=\pi/4$ ($\tau=0$) it is isotropic complex.

\subsection{Eigenvalue distribution of a single elliptic random matrix}

One can determine the eigenvalue distribution of $X$ using the same methods
as in Sec. III B. The only difference is that the propagators 
$\langle X_{ab}X_{cd} \rangle=
\langle X^\dagger_{ab} X^\dagger_{cd}\rangle$ (\ref{tau_prop}) do 
not vanish but are proportional to $\tau$. This leads to the following modification 
of the first Dyson-Schwinger equation (\ref{DSr1}):
\bq
\left( \begin{array}{cc} \sigma_{zz} & \sigma_{z\bar{z}} \\ 
\sigma_{\bar{z}z}& \sigma_{\bar{z}\bar{z}} \end{array} \right) =
\left( \begin{array}{cc} \tau g_{zz}  & g_{z\bar{z}} \\ 
g_{\bar{z}z} & \tau g_{\bar{z}\bar{z}} \end{array} \right)  ,
\label{DStau1}
\eq
while the second one (\ref{DSr2}) stays intact:
\bq
\left( \begin{array}{cc} g_{zz} & g_{z\bar{z}} \\ 
g_{\bar{z}z} & g_{\bar{z}\bar{z}} \end{array} \right) 
 = \left(\begin{array}{cc} z  - \sigma_{zz} & -\sigma_{z\bar{z}} \\ 
-\sigma_{\bar{z}z} & \bar{z}  - \sigma_{\bar{z}\bar{z}} \end{array} \right)^{-1}.
\label{DStau2}
\eq
These equations can be solved for $g_{zz}$. The solution reads 
\bq
g_{zz} = \left\{ \begin{array}{ll} 
\frac{\bar{z} - \tau z}{1-\tau^2}  
 & {\rm for} \ \frac{x^2}{(1+\tau)^2} + \frac{y^2}{(1-\tau)^2} \le 1,
\\ & \\
\frac{z - \sqrt{z^2 - 4\tau}}{2\tau} & {\rm otherwise}, \end{array}\right.
\eq
where $z=x+iy$. The non-holomorphic solution matches the 
holomorphic one on the ellipse. The eigenvalue density is \cite{SOMMERS}
\bq
\rho(z,\bar{z})= \frac{1}{\pi} \frac{\partial g_{zz}}{\partial \bar{z}} 
= \left\{ \begin{array}{ll}  \frac{1}{\pi(1-\tau^2)}  
 & {\rm for} \ \frac{x^2}{(1+\tau)^2} + \frac{y^2}{(1-\tau)^2} \le 1,
\\ & \\ 0 & {\rm otherwise}. \end{array}\right.
\eq
The parameter $\tau$ is a measure of flattening of the ellipse on which $\rho(z,\bar{z})>0$.
For $\tau=0$ the last equation reproduces the result for non-Hermitian complex matrices.
For $\tau \rightarrow 1$, the ellipse reduces to a  cut 
on the real axis. In order to determine the eigenvalue density in this
case one should first project the density for $\tau < 1$ onto the real axis:   
$\rho_*(x) = \int dy \rho(x,y)$, and then  take the limit
$\tau \rightarrow 1$.  One recovers the Wigner semicircle law 
$\rho_*(x) = \frac{1}{2\pi}\sqrt{4-x^2}$, as expected. 

\subsection{Eigenvalue distribution of a product of two or more elliptic random matrices}

We are now interested in the eigenvalue density 
of the product (\ref{product}) 
where $X_\mu$'s are drawn from a Gaussian ensemble with the measure (\ref{Ptau}). 
We shall show that the result is again given by Eq.~(\ref{main}) and hence exhibits a large degree of universality:
it does not depend on $\tau$ and is exactly the same even if each of the matrices $X_\mu$ is drawn from a Gaussian
ensemble with a different flattening parameter $\tau_\mu$. We will derive (\ref{main}) for $\X=\X_1\X_2$ and then make a comment on the generalization to $M>2$. 

\begin{figure}
	\includegraphics*[width=12cm]{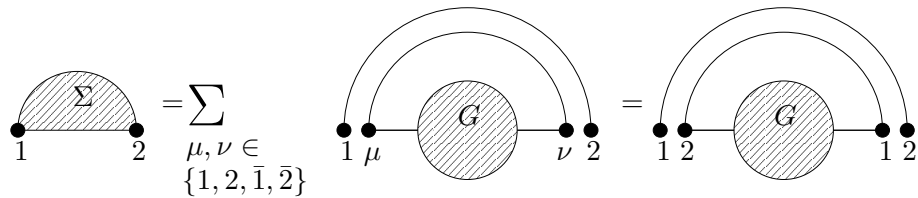}
	\caption{\label{diagram}Example of calculation of $\sigma_{12}$ in Eq.~(\ref{fds}). We write the second Dyson-Schwinger equation for $\Sigma_{12}$. The only non-vanishing propagator is the one between indices $1,2$ and $1,2$. Taking the trace of both sides of the equation we arrive at $\sigma_{12} = \tau_1 g_{21}$.
	}
\end{figure}

We will use the linearization and calculate first the eigenvalue density
of the matrix $Y$ (\ref{Y}) constructed from $X_1$ and $X_2$,
having the only non-vanishing propagators given by (\ref{tau_prop}) with two parameters 
$\tau_1$ and $\tau_2$. As before, first we have to determine 
the propagator structure for the block matrix ${\cal Y}$ (\ref{calY}) 
and then apply it to derive the Dyson-Schwinger equation. The matrix ${\cal Y}$ reads
\bq
{\cal Y} = \left(\begin{array}{cc} Y & 0 \\ 0 & Y^\dagger \end{array}\right) =
\left(\begin{array}{cccc} 0    & X_1 & 0 & 0 \\ 
                                  X_2 & 0    & 0 & 0\\
                              0 & 0 & 0 & X_2^\dagger \\ 
                                  0    & 0    & X_1^\dagger & 0 \end{array} \right)  .
\eq
The first non-vanishing propagator comes from the correlations between $X_\mu$'s and  
$X_\mu^\dagger$'s, exactly as in Eq.~(\ref{YYshort}):
\bq
\langle {\cal Y}_{12} {\cal Y}_{\bar{2}\bar{1}} \rangle = 
\langle {\cal Y}_{21} {\cal Y}_{\bar{1}\bar{2}} \rangle = \mathbbm{T}.
\eq
The next one comes from autocorrelations of $X_\mu$'s (\ref{tau_prop}) which are 
proportional to $\tau$,
\bq
\langle {\cal Y}_{12} {\cal Y}_{12} \rangle = \tau_1 \mathbbm{T} , \quad 
\langle {\cal Y}_{21} {\cal Y}_{21} \rangle = \tau_2 \mathbbm{T},
\eq
and the last one from autocorrelations of $X_\mu^\dagger$'s
\bq
\langle {\cal Y}_{\bar{1}\bar{2}} {\cal Y}_{\bar{1}\bar{2}} \rangle = \tau_1 \mathbbm{T} , \quad 
\langle {\cal Y}_{\bar{2}\bar{1}} {\cal Y}_{\bar{2}\bar{1}} \rangle = \tau_2 \mathbbm{T}.
\eq
Here $\mathbbm{T}$ denotes again a tensor with elements 
$T_{abcd} =\frac{1}{N} \delta_{ad} \delta_{bc}$, where $a,b$ are indices of the
first matrix and $c,d$ of the second one on the right-hand sides of the above equations.
All other correlations between the blocks of ${\cal Y}$ vanish.
We can now write two Dyson-Schwinger equations: 
\bq
\left( \begin{array}{cccc} \sigma_{11} & \sigma_{12} & \sigma_{1\bar{1}} & \sigma_{1\bar{2}} \\
                                   \sigma_{21} & \sigma_{22} & \sigma_{2\bar{1}} & \sigma_{2\bar{2}} \\
                                    \sigma_{\bar{1}1} & \sigma_{\bar{1}2} & \sigma_{\bar{1}\bar{1}} & 
                                    \sigma_{\bar{1}\bar{2}} \\
                                   \sigma_{\bar{2}1} & \sigma_{\bar{2}2} & \sigma_{\bar{2}\bar{1}} & 
                                   \sigma_{\bar{2}\bar{2}} \end{array} \right) 
=
\left( \begin{array}{cccc} 0                   & \tau_1 g_{21}  & g_{2\bar{2}} & 0 \\
                                   \tau_2 g_{12} & 0                    &  0                 & g_{1\bar{1}} \\
                                   g_{\bar{2}2}   & 0                    &  0                 & \tau_1 g_{\bar{2}\bar{1}} \\
                                   0                   & g_{\bar{1}1}    &  \tau_2 g_{\bar{1}\bar{2}} & 0 \end{array} \right) ,
\label{fds}
\eq
and 
\bq
\left( \begin{array}{cccc} g_{11} & g_{12} & g_{1\bar{1}} & g_{1\bar{2}} \\
													 g_{21} & g_{22} & g_{2\bar{1}} & g_{2\bar{2}} \\
													 g_{\bar{1}1} & g_{\bar{1}2}& g_{\bar{1}\bar{1}} & g_{\bar{1}\bar{2}} \\
                           g_{\bar{2}1} & g_{\bar{2}2}& g_{\bar{2}\bar{1}} & g_{\bar{2}\bar{2}} \\
       \end{array} \right) =
\left( \begin{array}{cccc} w-\sigma_{11} & -\sigma_{12} & -\sigma_{1\bar{1}} & -\sigma_{1\bar{2}} \\
                                   -\sigma_{21} & w-\sigma_{22} & -\sigma_{2\bar{1}} & -\sigma_{2\bar{2}} \\
                                    -\sigma_{\bar{1}1} & -\sigma_{\bar{1}2} & \bar{w}-\sigma_{\bar{1}\bar{1}} & 
                                    -\sigma_{\bar{1}\bar{2}} \\
                                   -\sigma_{\bar{2}1} & -\sigma_{\bar{2}2} & -\sigma_{\bar{2}\bar{1}} & 
                                   \bar{w}-\sigma_{\bar{2}\bar{2}} \end{array} \right)^{-1} .
\eq
In the first equation the off-diagonal blocks are the same as in the
previous section (\ref{Sg}). The diagonal blocks $\sigma_{ww},\sigma_{\bar{w}\bar{w}}$ now depend on $\tau_\mu$'s.
As an illustration we show in Fig.~\ref{diagram} a graphical representation 
of the equation for $\sigma_{12} = \tau_1 g_{21}$ which explains the flip of indices. 
Let us first look for a holomorphic solution, so assume that off-diagonal blocks of $g$ vanish:
$g_{w\bar{w}}=g_{\bar{w}w}=0$. In this case the above equations reduce to
\bq
\left( \begin{array}{cc} \sigma_{11} & \sigma_{12} \\
                                   \sigma_{21} & \sigma_{22}  \end{array} \right) =
                                   \left( \begin{array}{cc} 0              & \tau_1 g_{21}  \\
                                   \tau_2 g_{12} & 0     \end{array} \right) \quad , \qquad
\left( \begin{array}{cc} g_{11} &  g_{12} \\
                                  g_{21} & g_{22} \end{array} \right) =
\left( \begin{array}{cc} w - \sigma_{11} & -\sigma_{12} \\
                                   -\sigma_{21} & w  -\sigma_{22} 
                                   \end{array} \right)^{-1} ,
\eq
and the corresponding equations for $\sigma_{\bar{w}\bar{w}}$ and $g_{\bar{w}\bar{w}}$
being complex conjugate of those above. This gives
\bq
\left( \begin{array}{cc} g_{11} &  g_{12} \\
                                  g_{21} & g_{22} \end{array} \right) =
\left( \begin{array}{cc} w  &  -\tau_1 g_{21}\\
                                   -\tau_2 g_{12} & w   
                                   \end{array} \right)^{-1} ,
\eq
which has two solutions: one with $g_{11} = \frac{1}{w}$ and the other one with
$g_{11} = w/\sqrt{\tau_1\tau_2}$. We take the first one because it has the 
correct asymptotic behavior for large $w$. For this solution we have $g_{22}=\frac{1}{w}$ and 
$g_{12}=g_{21}=0$. The holomorphic solution has to be sewed with the non-holomorphic
one so that at the boundary $g_{12}=g_{21}=0$. If we assume that these elements vanish also 
inside the non-holomorphic region (and correspondingly $g_{\bar{1}\bar{2}}=g_{\bar{2}\bar{1}}=0$),
then the equation (\ref{fds}) reduces to
\bq
\left( \begin{array}{cccc} \sigma_{11} & \sigma_{12} & \sigma_{1\bar{1}} & \sigma_{1\bar{2}} \\
                                   \sigma_{21} & \sigma_{22} & \sigma_{2\bar{1}} & \sigma_{2\bar{2}} \\
                                    \sigma_{\bar{1}1} & \sigma_{\bar{1}2} & \sigma_{\bar{1}\bar{1}} & 
                                    \sigma_{\bar{1}\bar{2}} \\
                                   \sigma_{\bar{2}1} & \sigma_{\bar{2}2} & \sigma_{\bar{2}\bar{1}} & 
                                   \sigma_{\bar{2}\bar{2}} \end{array} \right) 
=
\left( \begin{array}{cccc} 0                   & 0  & g_{2\bar{2}} & 0 \\
                                     0  & 0                    &  0                 & g_{1\bar{1}} \\
                                   g_{\bar{2}2}   & 0                    &  0                 & 0 \\
                                   0                   & g_{\bar{1}1}    &  0 & 0 \end{array} \right) ,
\label{fds2}
\eq
with vanishing diagonal blocks. This equation is identical to the equation
with $\tau_1=\tau_2=0$ and was discussed in the previous section.
As we know it gives Girko-Ginibre distribution for the matrix $Y$ and hence we obtain (\ref{main}) for $X=X_1 X_2$.

One can repeat the whole reasoning for a product of more than two matrices.
One finds again that the solution $1/w$ valid outside the non-holomorphic region corresponds 
to vanishing blocks $g_{\mu\nu}=g_{\bar{\mu}\bar{\nu}}=0$ for 
$\mu \ne \nu$ and that it can be sewed with the non-holomorphic solution
for which the blocks also vanish. This gives $\sigma_{ww}=\sigma_{\bar{w}\bar{w}}$ 
and one obtains exactly the same equations as for $\tau_1=\dots=\tau_M=0$. 
Therefore, for $M>2$ the eigenvalue distribution of $Y$ is also given by the Girko-Ginibre law. 
This result is universal: the spectrum of $\X$ is given by Eq.~(\ref{main}) 
independently of whether we multiply two Hermitian 
matrices, or Hermitian by generic complex, or Hermitian by anti-Hermitian etc.
The limiting spectrum is always the same and differs only
by finite-size effects.

One can also extend this result to purely real matrices 
generated from the ensemble with a measure \cite{SOMMERS}
\bq
d\mu(\X) \sim \exp \left\{ -\frac{N}{2} \frac{1}{1-\tau^2} \left( 
{\rm Tr} XX^T - \tau  {\rm Tr} XX
\right)\right\} 
\prod_{ij} {\rm d} X_{ij}.
\label{Ptau2}
\eq
The case $\tau =1$ corresponds to symmetric real matrices, $\tau=-1$ to antisymmetric ones, and $\tau=0$ to isotropic real matrices.
The diagrammatic equations in the limit $N\to\infty$ are exactly the same as before, because the propagators have the same structure.

\section{Projection of the spectrum of a commutator of GUE matrices}
In this section we show that the conjecture made in \cite{biely} is not true. Let us consider a matrix $\X=\X_1\X_2$ which is a product of two Hermitian GUE matrices $\X_1,\X_2$. According to the formula (\ref{main}), the eigenvalue density of $\X$ is $\rho_\X(z,\bar{z}) = \frac{1}{2\pi |z|}$ for $|z|<1$ and zero otherwise. The projection of this function on the real (or imaginary) axis
gives 
\bq
\rho_*(x) = \frac{1}{\pi}\ln \frac{1+\sqrt{1-x^2}}{|x|}, \label{rho_proj}
\eq 
for $-1\leq x\leq 1$.
According to \cite{biely}, this result should be equal to the eigenvalue density $\rho_+(x)$
of $(\X_1\X_2 + \X_2^\dagger \X_1^\dagger)/\sqrt{8}$ or $\rho_-(x)$ of $i(\X_1\X_2-\X_2^\dagger \X_1^\dagger)/\sqrt{8}$. Up to a scaling factor $\sqrt{8}$, 
these spectral densities are equal to the spectra of the anticommutator $\left\{\X_1,\X_2\right\}$ or the commutator $i[\X_1,\X_2]$, because $\X_1=\X_1^\dagger,\X_2=\X_2^\dagger$.
Moreover, $\rho_-(x)=\rho_+(x)$ as follows from the observation that in the limit $N\to\infty$ all the moments of the commutator and the anticommutator are the same: $\Tr \left\langle [\X_1,\X_2]^k \right\rangle = \Tr \left\langle \{\X_1,\X_2\}^k \right\rangle$ for all $k=1,2,\dots$.

\begin{figure}
\includegraphics[width=7cm]{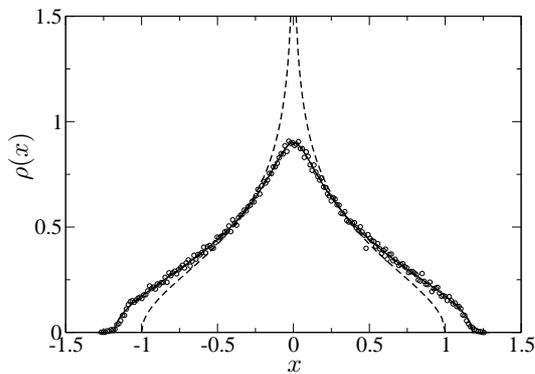}
\caption{\label{fig4}Comparison between $\rho_+(x)$ from Eq.~(\ref{rhox2y2}) (solid line), $\rho_*(x)$ from Eq.~(\ref{rho_proj}) (dashed line), and numerical simulations (circles) for $N=100$ (1000 matrices were generated).}
\end{figure}

We calculate now the eigenvalue density $\rho_+(x)$
of the rescaled anticommutator $\{\X_1,\X_2\}/\sqrt{8}$.
We define two matrices $\A=(\X_1+\X_2)/\sqrt{2}$ and $\B=(\X_1-\X_2)/\sqrt{2}$ which are
also mutually independent Hermitian matrices with a factorized probability measure
\bq
d\mu(\A,\B) \propto e^{-N/2 {\rm Tr} \A^2} e^{-N/2 {\rm Tr} \B^2} D\A D\B.
\eq
We have $\{\X_1,\X_2\}=\A^2-\B^2$. One can use the technique of free random variables \cite{SPEICHER}
to calculate the eigenvalue density of $\A^2-\B^2$ since in the limit $N\rightarrow \infty$ the
matrices $\A^2$ and $\B^2$ represent free random variables. The addition law for a sum of free variables is expressed in terms
of an $R$-transform or equivalently in terms of a Blue's function $B(z)$ which is
a functional inverse of the Green's function $G(B(z))=z$ and takes a simple 
form $B_{a+b}(z) = B_a(z) + B_b(z) - z^{-1}$, where $a$ and $b$ are free
random variables. In our case $a=\A^2$, $b=-\B^2$. The Green's function $G_a$ of $\A^2$ is a special case of the Green's function for Wishart distribution, while $G_b$ for $-\B^2$ corresponds to a reflected Wishart spectrum $\lambda\to -\lambda$, and hence
\bq
G_{a}(z) = \frac{1-\sqrt{1-4/z}}{2}, \;\;\;\; G_{b}(z) = -G_a(-z) = \frac{-1+\sqrt{1+4/z}}{2}.
\eq
The Blue functions for both cases read
\bq
B_{a}(z) = \frac{1}{z(1-z)}, \;\;\;\; B_{b}(z) = \frac{1}{z(1+z)},
\eq
and thus
\bq
B_{a+b}(z) = B_{a} + B_{b} -\frac{1}{z} = \frac{1+z^2}{z(1-z^2)}.
\eq
This equation has to be inverted for $G_{a+b}(z)$ which is the Green's function for the anticommutator:
\bq
z = \frac{1+G_{a+b}(z)^2}{G_{a+b}(z)(1-G_{a+b}(z)^2)},
\eq
which leads to a cubic equation for $G_{a+b}(z)$. The solution which has the correct behavior $G_{a+b}(z)\to 1/z$ for large $z$ reads
\bq
G_{a+b}(z) = \frac{1 + 3 z^2 + (-1 - 18 z^2 + 3 \sqrt{3}\sqrt{z^2 + 11 z^4 - z^6})^{2/3}}{3z(-1 - 18 z^2 + 3\sqrt{3}\sqrt{z^2 + 11 z^4 - z^6})^{1/3}}. \label{gx2y2}
\eq
Taking into account the scaling factor $\sqrt{8}$ we finally arrive at
\bq
	\rho_{+}(x) = -\frac{\sqrt{8}}{\pi} \mbox{Im} G_{a+b}(x\sqrt{8}+i0+) 
	= \frac{\sqrt{3}}{6\pi} \frac{1+24x^2-\left(1+144x^2-6\sqrt{6}\sqrt{x^2+88x^4-64x^6}\right)^{2/3}}{|x|\left(1+144x^2-6\sqrt{6}\sqrt{x^2+88x^4-64x^6}\right)^{1/3}}.
	\label{rhox2y2}
\eq
This is different from $\rho_*(x)$ from Eq.~(\ref{rho_proj}). In Fig. \ref{fig4} we compare both spectral densities and show also results
of numerical simulations which perfectly agree with (\ref{rhox2y2}). This falsifies the conjecture that if the spectrum of a non-Hermitian matrix is rotationally symmetric, it can be found by solving the symmetrized or antisymmetrized Hermitian problem.

\section{Conclusions}
The main result of this paper is that the eigenvalue density of a product of large, centered (with zero mean) Gaussian matrices assumes a very universal form (\ref{main}) with a single scaling parameter $\sigma$ representing the radius of a circular support in the complex plane and related to the amplitude of fluctuations of matrix entries. The matrices in the product
do not have to be identical and each of them may belong to a different elliptic ensemble. 

\begin{figure}
	\includegraphics*[width=12cm]{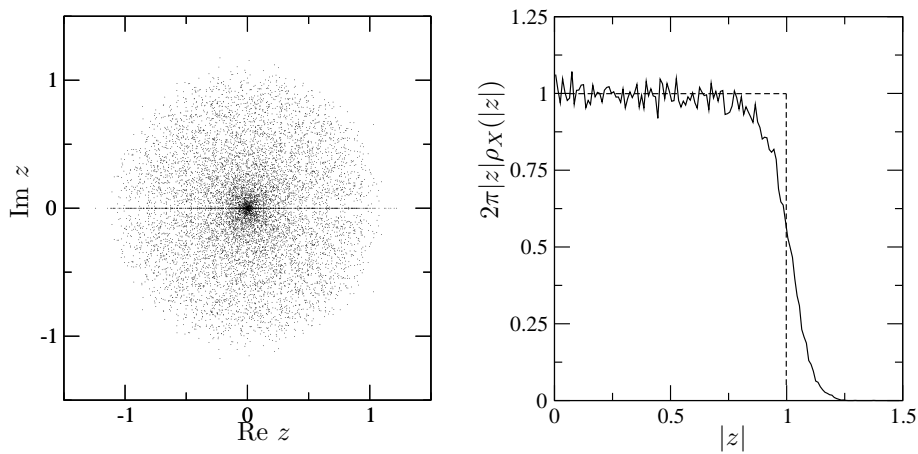}
	\caption{\label{fig5}Plots of numerically obtained $\rho_\X(|z|)$ for $\X_1,\X_2$ being two symmetric matrices which entries (upper triangle) are taken from uniform distribution $[-\sqrt{3/N},\sqrt{3/N}]$, for $N=200$ and for $1000$ matrices generated. Dashed line shows the theoretical distribution in the limit $N\to\infty$.}
\end{figure}

Taking into account the universality of the Wigner's semicircle law or the 
Girko-Ginibre distribution for matrices having their entries drawn from 
independent distributions, it is tempting to conjecture that our result 
will also hold in this setting. Namely, we suppose that the same asymptotic result 
holds for products of Wigner matrices having independent elements drawn from any centered distribution which fulfills Pastur-Lindeberg's condition \cite{PASTUR}.
To assess the validity of this conjecture we performed numerical simulations, assuming various distributions of elements of the matrices.
The only requirement was that the variance of the distribution was equal to $1/N$. We did not observe any deviations from (\ref{main}) for short-tailed distributions. In Fig.~\ref{fig5} we show an example for a uniform distribution with zero mean and variance $1/N$.

As far as future projects are concerned, it would be interesting to generalize the discussion to the Gaussian symplectic ensemble \cite{A} and to study microscopic properties of eigenvalues of the product of various types of Gaussian matrices from different invariant ensembles \cite{O,A,APS}. It would also be interesting to analytically derive the formula for the eigenvalue distribution of the product of $M$ matrices of finite size $N$ (see Fig.~\ref{fig2} in the middle). For the Girko-Ginibre ensemble \cite{KS} it is
given by $\rho(z)\cong{\rm erfc}(\sqrt{2}(|z|-1)\sqrt{N})/(2\pi)$. We expect a qualitatively similar behavior also for the product of matrices.

The discussion presented in this paper holds for Gaussian matrices for which
the first moment has zero mean, $\langle {\rm Tr} X_\mu \rangle = 0$. It would be interesting to check how it changes when $\langle {\rm Tr} X_\mu \rangle \ne 0$. This could be the first step towards a generalization of Voiculescu's $S$-transform composition rule \cite{VOICULESCU} for calculating the eigenvalue density of asymptotically large matrices representing free random variables, to the case when their product has complex eigenvalues. 

\subsection*{Acknowledgements}
We thank the Polish Ministry of Science grants NN202~229137 (2009-2012) (ZB) and NN202~105136 (2009-2011) (RJ).
RJ was partially supported by the Marie Curie ToK KraGeoMP (SPB 189/6.PRUE/2007/7).
BW acknowledges partial support by the EC-RTN Network ENRAGE under grant No.~MRTN-CT-2004-005616 and EPSRC grant EP/030173.

\end{document}